\documentclass[9pt,twocolumn]{article}

\usepackage{cite}
\usepackage{graphicx}
\usepackage{epstopdf}
\usepackage{color}
\usepackage{amsmath}
\usepackage{mathptmx}
\usepackage{transparent}
\usepackage{authblk}
\graphicspath{{figures/}}

\title{\bf Micro lensing induced lineshapes in a single mode cold-atom hollow-core fiber interface}

\author[1,*]{Mohammad Noaman}
\author[1]{Maria Langbecker}
\author[1]{Patrick Windpassinger}

\affil[1]{EQOQI, QUANTUM, Institute of Physics, Johannes Gutenberg-University, 55099 Mainz }
\affil[*]{Corresponding author: monoaman@uni-mainz.de}

\begin{document}

\twocolumn[
\begin{@twocolumnfalse}
	\maketitle
	\begin{abstract}
 	\bf {We report on the observation of strong transmission line shape alterations in a cold-atom hollow-core fiber interface. We show that this can lead to a significant overestimation of the assigned resonant optical depth for high atom densities. By modeling light beam propagation in an inhomogeneous dispersive medium, we attribute the observations to micro lensing in the atomic ensemble in combination with the mode selection of the atom-fiber interface. The approach is confirmed by studies of Rydberg EIT line shapes.}\\
	\end{abstract}
\end{@twocolumnfalse}
]

Single mode light-matter interfaces are at the heart of quantum optics, quantum information, quantum communication and sensing. Besides Jaynes-Cummings type cavity systems and free space arrangements for neutral atoms \cite{Kimble2008, Kuhn1999}, Rydberg atoms \cite{Saffman2010, Pritchard2012}, ions \cite{Mundt2002} or artificial atoms like quantum dots \cite{Reithmaier2004}, hollow-core (HC) fibers into which the respective quantum systems are loaded have recently attracted considerable attention \cite{Bajcsy2009, Langbecker2017,Hilton2018,Xin2018}. Here, a single mode light beam is coupled to e.g. an atomic ensemble over an extended distance of several centimeters \cite{Bajscy2011}. This way, the Rayleigh length of a focused beam, which usually determines the length of the efficient coupling region, is overcome and the overall light-matter coupling strength is considerably enhanced. A typical figure of merit in this context is the resonant optical depth (OD) which has been shown to exceed a value of 1000 in a cold-atom filled hollow-core fiber interface \cite{Blatt2014}.

Micro lensing in an inhomogeneous, dense, near-resonant atomic ensemble \cite{Roof2015, Han2015} can significantly alter the shape of the traversing light beam \cite{Gilbert2018}. In the following, we discuss the impact of such a micro lensing effect on a single mode cold-atom hollow-core fiber interface. In particular, we demonstrate the impact of atomic micro lensing onto the observed atomic transition line shape, both for a dipole allowed E1 transition and a two-photon Rydberg electromagnetically induced transparency (EIT) system. We argue that our observations present an under-explored  source of line asymmetries in such single-mode light-matter interfaces which can significantly alter the assignment of ODs or atomic densities.

\begin{figure}[htbp]
	\centering
	\def\svgwidth{\linewidth}
	\begin{tiny}
		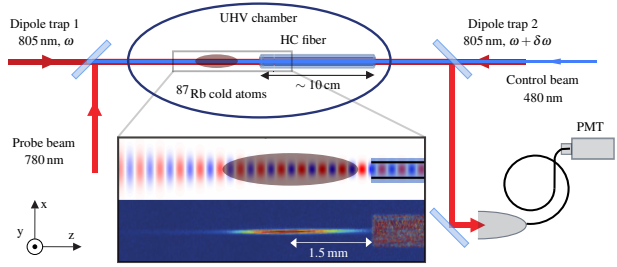
	\end{tiny}
	\caption{Schematic of the experimental setup. The main part shows a sketch of the vacuum system with the relevant beam paths. The inset shows a sketch of the potential profile of the movable optical lattice  (not to scale) and an absorption image of cold atoms close to the fiber tip.}
	\label{fig:exp_stp}
\end{figure}

For thoroughly studying the effect of micro lensing, we make use of our highly controlled cold-atom hollow-core fiber interface \cite{Langbecker2017, Langbecker2018}. A schematic of our experimental setup is depicted in Figure \ref{fig:exp_stp}, where we load about $10^5$ atoms of $^{87}$Rb from a magneto-optical trap into a movable optical lattice about 5\,mm away from the tip of a hollow-core photonic crystal fiber. The fiber has a $60\,\mu$m core and a mode field diameter of $\sim 42$\,$\mu$m with coupling efficiencies above 90\% \cite{Benabid2011}. By controlling the relative detuning of the optical lattice beams, we can precisely position the atoms in front or inside the fiber \cite{Langbecker2018, Okaba2014}. A typical absorption image of atoms transported towards the fiber is shown in Figure \ref{fig:exp_stp}. By coupling a (near-resonant) Gaussian shaped probe beam into the fundamental mode of the fiber, i.e. into the same mode in which the atoms are trapped and transported in, we complete our light-matter interface. For probing, we use the $5S_{1/2}(F=2) \leftrightarrow 5P_{3/2}(F=3)$ E1 transition of $^{87}$Rb.
After passing through the fiber, the probe beam is sent to a fiber coupled photo multiplier tube (PMT). In a typical experimental sequence, the atoms are probed continuously with $\sim 200$\,pW probe power after transport and release from the optical lattice by recording the transmitted signal. The background corrected data contain the temporal absorption profile due to the expansion of the cloud, i.e. we sample different atomic density distributions. To obtain the probe frequency dependent absorption line shape, we repeat the experiment for various probe detunings. For Rydberg EIT, we apply a 480\,nm control beam, stabilized to the $5P_{3/2}(F=3) \leftrightarrow 29 S_{1/2}$ transition, simultaneous to the probe beam \cite{Langbecker2017}. This EIT control beam is coupled to the fundamental mode of the fiber as well and counter-propagates the probe beam. During the probing, micro lensing effects change the shape of the beam, i.e. its mode composition and thus the coupling and guiding of the probe light through the hollow-core fiber and the PMT fiber. This micro lensing, and consequently the guiding of the fiber, is atom density and probe frequency dependent and thereby impacts the observed absorption line shapes. The core of this manuscript is to gain detailed understanding of light propagation in such atomic media and the observed line shapes.

Before discussing the experimental results, we set up a theoretical model. The model is based on the paraxial wave equation (PWE) with a spatially varying refractive index according to the atom density. We approximate the trapped atoms' density by a Gaussian distribution:
\begin{align}\label{eq:rho}
\rho &=  \rho_0 \exp \left\{ -\frac{z^2}{2\sigma_z^2} -\frac{x^2 + y^2}{2\sigma_r^2}  \right\},
\end{align}
where $\sigma_r$ and $\sigma_z$ correspond to the radial and axial size of the cloud. For a near-resonant weak probe, the optical response of a 2-level atom is imprinted in the electrical susceptibility \cite{Roof2015}:
\begin{align}\label{eq:sus}
\chi &=  - \frac {\sigma_0 \rho(x,y)}{k_0} \frac{2\delta_p/\gamma -i }{1+4(\delta_p/\gamma)^2},
\end{align}
where $\sigma_0$ is the resonant scattering cross-section, $\delta_p$ is the probe detuning from the resonance and $\gamma$ is the natural linewidth of the two level system. $\chi$ gives rise to the real and the imaginary part of the refractive index which in turn lead to dispersion $n_r$ and absorption $n_i$ as
\begin{align}\label{eq:ref_indx}
n_r &=  1 + \frac{1}{2}Re(\chi), \\
n_i &=  \frac{1}{2}Im(\chi).
\end{align}

\begin{figure}[htbp]
	\centering
	\def\svgwidth{\linewidth}
	\begin{tiny}
		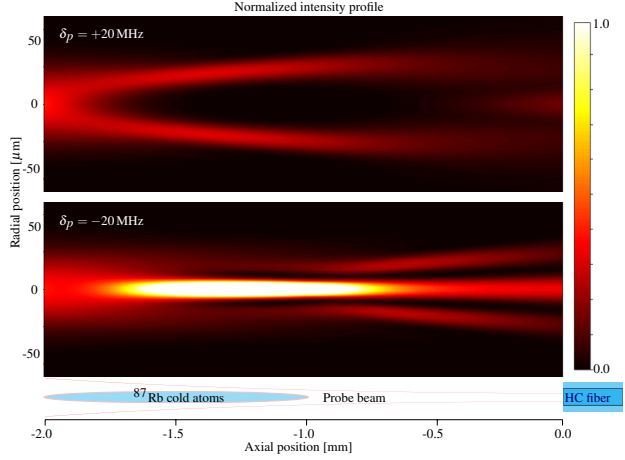
	\end{tiny}
	\caption{Micro lensing induced intensity profiles for positive and negative detunings of the probe beam. The schematic indicates the position of the atomic cloud and fiber as used in the calculations.}
	\label{fig:int_profile}
\end{figure}

In our highly mode matched and tightly confined light-matter interface, the spatially varying susceptibility caused by the atomic density distribution has a significant impact on the light propagation. The density variation is very prominent in the radial direction, while it can be considered constant along the cloud axis for the characteristic cloud length.
Under the slowly-varying-envelope-approximation (SVEA), the complex amplitude of the electric field is described by \cite{Roof2015}
\begin{align}\label{eq:svea}
\frac{\partial \boldsymbol{A}}{\partial z} = \frac{i}{2k}  \nabla_\perp^2 \boldsymbol{A} + \frac{ik}{2n_r^2} \chi \boldsymbol{A},
\end{align}
where $\boldsymbol{A}$ is the complex electric field and $k = 2\pi/\lambda$ is the wave vector. We numerically solve Equation \ref{eq:svea} using the so-called "split-step" algorithm \cite{Weideman1986}, where the atomic cloud is divided into thin slices, through which the light field propagation is calculated. Consequentiality, for our typical experimental parameters and two selected detunings, the obtained intensity profile in the x-z plane through the center of the cloud is shown in Figure \ref{fig:int_profile}. The effect of micro lensing induced  focusing and defocusing can clearly be observed.

\begin{figure}[htbp]
	\centering
	\def\svgwidth{\linewidth}
	\begin{tiny}
		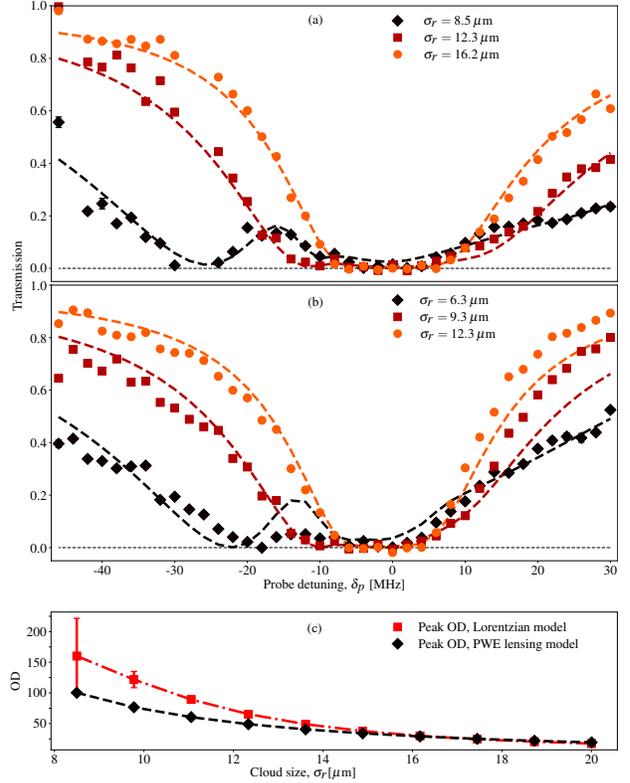
	\end{tiny}
	\caption{Transmission profiles for the atoms (a) outside and (b) inside the fiber. Markers and the dashed lines represent the data with statistical error and the result of the model. (c) Peak OD extracted from the Lorentzian line model and the PWE lensing model.}
	\label{fig:od_fit}
\end{figure}

To further include the guiding and mode selection properties of the HC fiber and the PMT detection, we estimate the transfer efficiency $\eta$ with the overlap integral of the electric field after passing though the cloud truncated by the HC fiber wall $\mathcal{E}_1$ and the unaltered field without atoms $\mathcal{E}_0$:
\begin{align}\label{eq:eta}
	\eta = \frac {|\int \mathcal{E}_1^* \mathcal{E}_0 dA |^2}{\int |\mathcal{E}_1|^2 dA \int |\mathcal{E}_0|^2 dA}.
\end{align}

To experimentally test our model, we exploit the high degree of control over the position of the atoms along the fiber axis.  We first consider ensembles $\sim 1.5$\,mm in front of the fiber, as shown in Figures \ref{fig:exp_stp} and \ref{fig:int_profile}. To vary the probed atomic density distribution, we exploit the natural kinetic expansion of the cloud and we evaluate the normalized transmission at different times after releasing the atoms from the optical lattice.  Figure \ref{fig:od_fit}(a) shows the observed absorption line shapes for three different atom densities (i.e. $\sigma_r$) together with the result from the theoretical model. The estimated evolution of the cloud size $\sigma_r$ corresponds to the expansion expected from the typical temperature of the atoms \cite{Langbecker2018}.

As can clearly be seen in Figure \ref{fig:od_fit}(a), for small cloud sizes, i.e. for high atomic densities, the absorption profile develops a significant asymmetry with an additional off-resonant peak near $-15$\,MHz detuning. This is well explained by the micro lensing discussed in our model as the simulation fits very well to the data. Obviously, the observed effects cannot be covered by a simple Lorentzian line shape model for the transmitted signal $T=~\exp[-OD/(1+~4(\delta_p/\gamma)^2)]$. The asymmetries are also not accounted for by effects of neighboring transitions. To demonstrate this, we on the one hand calculate the peak OD from the different $\sigma_r$ obtained by our PWE solution and on the other hand extract the OD from the simple Lorentzian line fit to the data.  Both results are plotted in Figure \ref{fig:od_fit}(c). It is evident that for cloud sizes $\sigma_r<14\,\mu$m, a simple model would overestimate the OD significantly. The large fitting error also confirms that the line shapes cannot be accounted for in a simple model. 

These observations remain generally valid for atoms transported into the HC fiber as shown in Figure \ref{fig:od_fit}(b). Again, our simple micro lensing model covers very well the observed asymmetries and shifts in the absorption profile. We attribute the remaining small deviations to the extremely simple modelling of the guiding and mode selection properties of the HC fiber. 

\begin{figure}[htbp]
	\centering
	\def\svgwidth{\linewidth}
	\begin{tiny}
		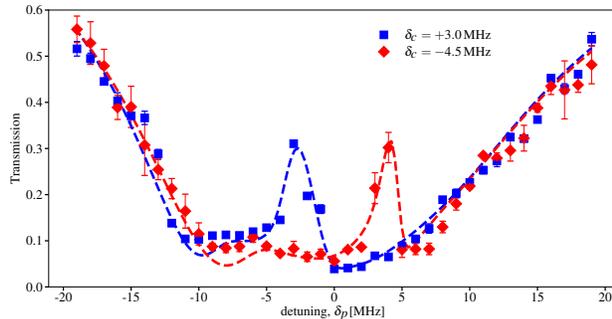
	\end{tiny}
	\caption{Experimental data (markers) and theory model results (dashed lines) for Rydberg EIT in front of the HC fiber. The error bars represent the statistical error.}
	\label{fig:eit_fit}
\end{figure}

As is evident from the previous discussion, the atomic ensemble induced dispersion in the probe beam has a crucial impact on the observed line shapes in single mode light-matter interfaces, especially when the ensemble size is comparable to the probe beam size. In this connection, EIT processes are known for manipulating optical properties in a very controlled way \cite{Fleischhauer2005}. On the one hand, EIT offers a flexible means to control the dispersion. On the other hand, due to its application in quantum optics, quantum communication and sensing, the observed EIT line shapes play a crucial role in the interpretation of the underlying physics. We therefore study the impact of micro lensing on the Rydberg EIT signals close to the HC fiber. To avoid any line shift and broadening due to the electric fields present at the fiber surface we keep the atoms outside the fiber. In EIT, the susceptibility $\chi_{EIT}$ is expressed as \cite{Gea1995}:
\begin{align}\label{eq:eit_sus}
\chi_{EIT} =   \frac {\sigma_0 \rho(x,y)}{k_0} \frac{i\gamma/2}{\gamma/2-i\delta_p + \frac{\Omega_c^2(x,y)}{\Gamma/2 -i(\delta_p+\delta_c)}},
\end{align}
where $\Omega_c(x,y)$ is the EIT control Rabi frequency of the $5P_{3/2}(F=3) \leftrightarrow 29 S_{1/2}$ transition with Gaussian distribution, $\delta_c$ is the control beam detuning from the Rydberg state and $\Gamma$ is the decoherence rate of the Rydberg state. By using this modified susceptibility in the numerical calculations where we allow $\Gamma$, $\Omega_c$ and $\delta_c$ to vary, we obtain the theory curves plotted in Figure~\ref{fig:eit_fit} together with the corresponding experimental data. We show sample data and model results for two different detunings for the EIT control laser frequency with respect to the Rydberg transition. The model reproduces the observed asymmetries in the overall absorption profile and, in particular, the asymmetries in the EIT line very well. Here the values of $\chi_{EIT}$ remain similar to that of E1 transition for $\delta_p$ away from two-photon transitions which produce similar asymmetric line shapes as shown in Figure \ref{fig:od_fit}, while the exact line shapes near the two photon transition, i.e. $\delta_p = -\delta_c$, result from the expected off-resonant EIT in combination with the lensing effect. Generally, asymmetries in EIT line shapes are attributed to Rydberg interactions or super radiance \cite{Firstenberg2016, Melo2016, Weller2016, Weatherill2008, Langbecker2017, Keaveney2012} but our model indicates that micro lensing has a considerable effect. With the chosen Rydberg state, $29S_{1/2}$, we also minimize the possibility of significant interactions which makes our technique suitable for the line shape studies. It would be a topic of further investigation to include interactions by e.g. exciting atoms to very high Rydberg states, however this is beyond the scope of the work presented here.

In conclusion, based on the numerical simulation of a paraxial Helmholtz equation with a spatially dependent refractive and absorptive medium, we have developed a model for micro lensing in a single mode hollow-core fiber cold-atom interface. This model reproduces the absorption line shapes observed, both for standard E1 transitions and Rydberg EIT configurations. The findings have a critical impact on the figure of merit of such single mode light-matter interfaces -- the optical depth -- as lensing can lead to artificially raised or lower transmission values for different probe frequencies.\\

We thank Ronja Wirtz for help with the experiment and Parvez Islam and Wei Li for useful discussions. The work has been supported by the DFG SPP 1929 GiRyd and FP7-PEOPLE-2012-ITN-317485 (QTea). 
\bigskip
\bibliography{mybib}{}
\bibliographystyle{plain}

\end{document}